\documentclass[aps,prb,preprint,unsortedaddress,showpacs]{revtex4}
\usepackage{amssymb}
\usepackage{graphicx}

\begin{document}

\title{Heat Capacity of PbS: Isotope Effects}
\author{M. Cardona}
\email[Corresponding author:~E-mail~]{M.Cardona@fkf.mpg.de}
\author{R. K. Kremer}
\author{R. Lauck}
\author{G. Siegle}
\affiliation{Max-Planck-Institut f{\"u}r Festk{\"o}rperforschung,
Heisenbergstr. 1, D-70569 Stuttgart, Germany}

\author{J.Serrano}
\affiliation{ European Synchrotron Radiation Facility, Bo\^{\i}te
Postale 220, 38043 Grenoble, France}
\author{A.H. Romero}
\affiliation{CINVESTAV, Departamento de Materiales, Unidad
Quer$\acute{e}$taro, Quer$\acute{e}$taro, 76230, Mexico}

\date{\today}

\begin{abstract}
In recent years, the availability of highly pure stable isotopes has
made possible the investigation of the dependence of the physical
properties of crystals, in particular semiconductors, on their
isotopic composition. Following the investigation of the specific
heat ($C_p$, $C_v$) of monatomic crystals such as diamond, silicon,
and germanium, similar investigations have been undertaken for  the
tetrahedral diatomic systems ZnO and GaN (wurtzite structure), for
which the effect of the mass of the cation differs from that of the
anion. In this article we present measurements for a semiconductor
with rock salt structure, namely lead sulfide. Because of the large
difference in the atomic mass of both constituents ($M_{\rm Pb}$=
207.21 and  ($M_{\rm S}$=32.06 a.m.u., for the natural isotopic
abundance) the effects of varying the cation and that of the anion
mass are very different for this canonical semiconductor. We compare
the measured temperature dependence of $C_p \approx C_v$, and the
corresponding derivatives with respect to ($M_{\rm Pb}$ and $M_{\rm
S}$), with \textit{\textit{ab initio}} calculations based on the
lattice dynamics obtained from the local density approximation (LDA)
electronic band structure. Quantitative deviations between theory
and experiment are attributed to the absence of spin-orbit
interaction in the ABINIT program used for the electronic band
structure calculations.
\end{abstract}

 \maketitle


 \email{M.Cardona@fkf.mpg.de}

\section{Introduction}

Lead sulfide belongs to the family of the lead chalcogenides (PbS,
PbSe, PbTe) which are small bandgap semiconductors that crystallize
in the rock salt structure. They differ from the nowadays more
commonly encountered tetrahedral semiconductors (e.g. GaAs, ZnO,
with zincblende and wurtzite structure, respectively) in that they
possess 10 valence electrons per formula unit, as opposed to 8 for
the zincblende- and wurtzite-type materials. The lead chalcogenides
are found as minerals in nature, bearing the names: galena (PbS),
clausthalite (PbSe) and altaite (PbTe). Whereas galena is rather
abundant and is found as large, highly perfect crystals,
clausthalite and altaite are rare. Some semiconducting properties of
PbS have been known since the mid 1800, in particular the rectifying
properties of metal-PbS contacts  which were used as detectors in
early radio receivers \cite{Braun}. The thermoelectric properties of
PbS were reported as early as 1865 (Ref. \onlinecite{Stefan}). The
photoresistive and photovoltaic properties were reported in 1930
(Ref. \onlinecite{Lange}). A number of reviews which summarize the
physical properties of PbS have appeared \cite{Ravich,Dalven}.
Particularly interesting is the anomalous dependence of the
electronic energy gap on temperature \cite{Paul}.

The specific heat of PbS was first measured by Eastman and Rodebush
above 63 K and subsequently by Parkinson and Quarrington  for
temperatures between 20 and 270K \cite{Eastman,Parkinson}. Such
measurements were extended down to 1K by Lykov and Chernik
\cite{Lykov}. Although the typical carrier concentrations of the PbS
samples were between 10$^{18}$ and 10$^{19}$ cm$^{-3}$, no evidence
of a free carrier contribution to the specific heat was found,
except possibly below 2K \cite{Keesom}. Thus the measured specific
heat represents the contribution of the lattice vibrations vs.
temperature, obtained at constant pressure ($C_p$). Most lattice
dynamical calculations, such as those of Elcombe, represent the
specific heat measured at constant volume ($C_v$) \cite{Elcombe}.
The difference between $C_p$ and  $C_v$  is given by \cite{Kremer}:

\begin{equation}
C_p(T) - C_v(T) = \alpha_v^2(T) \cdot B \cdot V_{{mol}} \cdot T,
 \label{equ1}
\end{equation}

where $\alpha_v$ is the temperature dependent thermal expansion
coefficient, $B$ the bulk modulus and $V_{mol}$  the molar volume.
According to Eq. (1), the difference $C_p$ - $C_v$ increases with
increasing $T$.  Replacing standard values of  $\alpha_v (T)$, $B$
and $V_{mol}$ found in the literature we obtain from Eq. (1) the
difference $C_p$ - $C_v$ = 0.18 J/mol K at the highest temperature
used in our measurements ($\sim$280 K). This difference is smaller
than typical error bars in our measurements and will be neglected
here. We shall therefore denote the specific heat, either
theoretical or experimental, by $C_v$ .

\section{Theoretical Details}

The specific heat of PbS vs. $T$ has been calculated in the harmonic
approximation from the vibrational free energy $F(T)$ as obtained
with the ABINIT code \cite{ABINIT}. This code derives the phonon
dispersion relations from total energy calculations and uses them to
evaluate $F(T)$ with the expression:

\begin{equation}
F(T)=-\int_0^\infty (\frac{\hbar\omega}{2} +  k_{B}T
ln[2n_B(\omega)]) \rho(\omega)d\omega\label{eq2}
\end{equation}

From Eq. (\ref{eq2}) the specific heat is obtained with:
\begin{equation}
C_v =-T\left(\frac{\partial^2 F}{\partial T^2}\right)_v.\label{eq3}
\end{equation}

In Eq. (\ref{eq2}), $k_{B}$ is the Boltzmann constant, $n_B$ the
Bose-Einstein factor, and $\rho(\omega)$ the phonon density of
states (PDOS). The high frequency cutoff of the latter defines the
upper limit of integration in Eq. (\ref{eq2}).

The theoretical calculations were performed in the framework of the
density functional theory (DFT) by using a local density
approximation for the exchange-correlation Hamiltonian as
implemented in the ABINIT package. Kohn-Sham orbitals are expanded
in plane waves with the use of the Hartwigsen-Goedecker-Hutter
pseudopotentials  to describe the valence electrons (4 electrons for
Pb and 6 electrons for S) \cite{Hartwigsen}. A 70 Ry cutoff was used
and a Monkhorst Pack grid of 8$\times$8$\times$8 points was used to
describe the electronic properties in the Brillouin zone.  Before
vibrational calculations were performed, the cell parameter was
optimized to a value of 5.808 $\rm{\AA}$ which was used thereafter.
The ABINIT program determines the elements of the dynamical matrix
by perturbation theory. They were calculated for a
12$\times$12$\times$ 12 mesh and then Fourier interpolated to obtain
the thermal properties discussed here.

The ABINIT program determines the elements of the dynamical matrix
by using perturbation theory.  The various isotopic masses used in
the calculation were introduced at the level of the dynamical
matrix, assuming that the matrix elements of the latter do not
depend on those masses. Possible dependence of dynamical matrix
elements (i.e. force constants) on mass can be estimated from Raman
and inelastic neutron scattering measurements \cite{Cardona}. They
are several orders of magnitude smaller than the corresponding
isotope mass changes and thus can be neglected for the purpose of
the present work.

Under these assumptions and for monatomic crystals (diamond, Ge, Si)
all frequencies $\omega$ are proportional to $M^{-0.5}$. Because of
the appearance of the argument ($\omega/T$) in $n_B$ (Eq. 2), it is
possible to relate the derivative of $C_v$ with respect to the
isotopic mass $M$ to that with respect to $T$. The following
expression is found \cite{Sanati}

\begin{equation}
\frac{d \ln (C_{p,v}/T^3)}{d \ln M} =  \frac{1}{2}\,\,(3 + \frac{d
\ln (C_{p,v}/T^3)}{d \ln T}). \label{eq4}
\end{equation}

In binary (or multinary) crystals, the masses of the different
elements have a different effect on each phonon frequency which, in
general, cannot be evaluated with simple arguments. This effect
depends on the phonon eigenvectors or, equivalently, on the DOS
projected on the atom under consideration \cite{Menendez}.  As we
shall see below, for PbS because of the large difference in the
masses of Pb and S, the lattice vibrations are predominantly Pb-like
up to 120 cm$^{-1}$ (acoustic phonons). Higher frequency vibrations
(optic phonons) are overwhelmingly S-like. The frequency of the
acoustic phonons thus should vary, to a good approximation, like the
inverse square root of the Pb mass ($M_{\rm Pb}^{-1/2}$) whereas
that of the optic phonons should vary like $M_{\rm S}^{-1/2}$.

For computational reasons (the limited number of
\textbf{\textit{k}}-points being sampled) the calculated specific
heats are not reliable for $T$ below $\sim$5K. A similar low $T$
limit also applies to the experimental data due to the small values
of $C_v$ and the limited accuracy of the experimental procedure
using sample masses of  $\sim$25mg for the isotope enriched samples.
$T<$ 5K corresponds to the Debye limit in which $C_v$ is
proportional to $(T/\theta^3)$ . This expression enables us to
derive for the mass derivatives in the limit $T\rightarrow$0 the
following analytic expressions \cite{Serrano}:

\begin{equation}
\frac{d \ln C_v/T^3}{d \ln M_{\rm Pb}} =  \frac{3}{2} \frac{M_{\rm
Pb}}{M_{\rm Pb} + M_{\rm S}} = 1.30\qquad\qquad \frac{d \ln
C_v/T^3}{d \ln M_{\rm S}}  =  \frac{3}{2} \frac{M_{\rm S}}{M_{\rm
Pb} + M_{\rm S}} = 0.20\label{eq5}
\end{equation}

The values given in Eqs. (\ref{eq5}) can be used to extrapolate
those obtained at 5K to $T \rightarrow$ 0.

In order to achieve convergence of the heat capacity and its
logarithmic derivatives with respect to $M_{\rm Pb}$ and $M_{\rm S}$
a grid of 80$\times$80$\times$80 \textbf{\textit{q}} points in the
irreducible Brillouin zone was used for the evaluation of the PDOS
and the thermodynamic properties. The number of \textbf{\textit{q}}
points used in the integrations (e.g., Eq. \ref{eq2}) is
particularly important for the convergence of the calculations of
the mass derivatives at low temperature. These calculations were
performed by evaluating the differences of $C_v$ for two different
lead and sulfur pairs of isotopes. Below 5K, the calculated
derivatives depended strongly on the pair of isotopes and the sample
chosen. We kept only the values for $T \rightarrow$ 0 as given by
Eq. \ref{eq5}

\section{Lattice Dynamics: Dispersion Relations and Density of Phonon States}

\begin{figure}[tbph]
\includegraphics[width=8cm ]{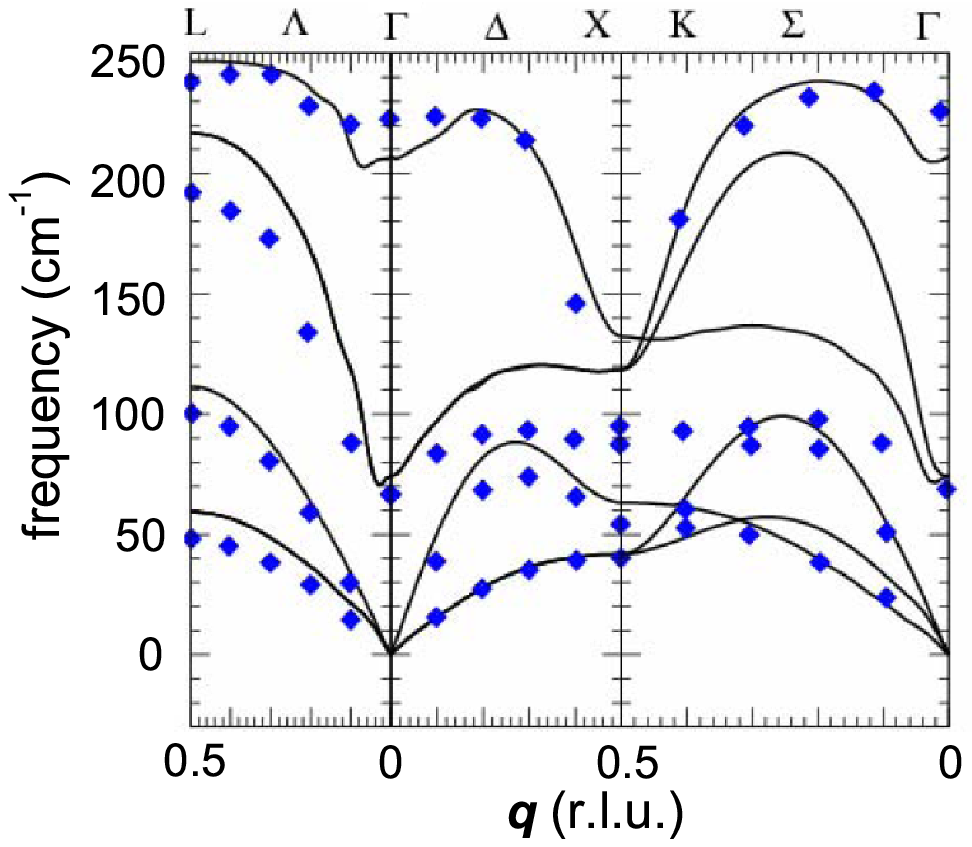}
\caption{(color online) Phonon dispersion relations of  PbS assuming
the natural isotope compositions for Pb and S obtained from the ab
initio electronic band structure using the ABINIT code (black solid
line). Experimental points obtained by inelastic neutron scattering
measurements are denoted by (blue) diamonds \cite{Elcombe}. Note
that the TO band is considerably higher than the experimental data
points around the X and  K points, a fact which is attributed in the
text to the lack of spin-orbit interaction in the ABINIT code
used\cite{ABINIT}.} \label{fig1}
\end{figure}

Figure \ref{fig1} displays the phonon dispersion relations of
natural PbS obtained from the \textit{ab initio} electronic band
structure using the procedure described above (solid lines).
Experimental points obtained by inelastic neutron scattering
(INS)\cite{Elcombe} are denoted by diamonds. Of particular interest
is the large LO - TO splitting which, by virtue of the
Lyddane-Sachs-Teller relation \cite{Lyddane}, indicates a large
value of the low-frequency dielectric constant ($\epsilon (0) \sim$
180, $\epsilon (\infty)\sim$ 18) which is related to the nearly
ferroelectric character of PbTe \cite{Grabecki}.

The INS points in Fig. \ref{fig1}  agree reasonably well with the
calculated dispersion relations except for the TO modes along the
$\Sigma$ and the $\Delta$ directions. Previous calculations were of
a semi-empirical nature with parameters adjusted to fit the INS
results \cite{Elcombe,Upanhyaya}. It is therefore not surprising
that the agreement between the calculated dispersion relations and
the INS results is somewhat better than that of Fig. \ref{fig1}.

We conjecture that the discrepancies between theory and experiments
displayed in Fig. \ref{fig1} may be due to the lack of the
spin-orbit interaction contribution in the corresponding LDA
Hamiltonian. This conjecture is supported by recent \textit{ab
initio} calculations of the phonon dispersion relations of bismuth
performed with and without spin-orbit interaction \cite{Diaz}.
\textit{ab initio} electronic band structure calculations which
include spin-orbit coupling have been recently performed for PbS,
PbSe, and PbTe  but, to our knowledge, they have not been used to
derive the phonon dispersion relations. \textit{Ab initio}
calculations of the LO and TO frequencies at \textbf{\textit{q}}=0
have been reported for PbTe \cite{Zein}.

The TO bands of Fig. \ref{fig1} are nearly dispersionless along
$\Sigma$ and $\Delta$. This fact suggests phonon modes in which the
lead and the sulfur vibrations are only weakly coupled, a conjecture
which is confirmed by the calculated phonon eigenvectors and also by
the PDOS displayed in Fig. \ref{fig2} as projected on the lead (
predominant below 120 cm$^{-1}$) and the sulfur (predominant above
120 cm$^{-1}$) eigenvectors. The van Hove singularities of the TA
modes at the L and X points are responsible for the sharp, Pb-like
spike seen at $\sim$50 cm$^{-1}$ (cf. Fig. \ref{fig2}), whereas the
flat regions of the TO bands are responsible for the much lower peak
found at $\sim$130 cm$^{-1}$.

\section{Experimental Procedure}
The preparation and some properties of the PbS samples are described
in Ref. \onlinecite{Sherwin}.  Two types of samples with the natural
isotopic composition ($M_{\rm Pb}$ =207.21 a.m.u., $M_{\rm S}$
=32.05 a.m.u. ) were used. One was a piece of natural galena (from
Creede, CO, USA,  16 mg) whereas the others were pieces of the same
size taken from the synthetic ingots used in Ref.
\onlinecite{Sherwin}. We also measured a piece of the
$^{\rm{nat}}$Pb$^{\rm{34}}$S of Ref. \onlinecite{Sherwin} and a
piece from an ingot prepared in the same way with the isotopic
composition $^{\rm{208}}$Pb$^{\rm{nat}}$S. All samples were p-type
(as determined from the sign of the thermoelectric power), the
natural sample had a carrier concentration of 2$\times$10$^{17}$
cm$^{-3}$ whereas the synthetic ones had carrier concentrations
around 10$^{18}$ cm$^{-3}$ (as determined from the infrared plasma
edge, see Ref. \onlinecite{Sherwin}).

The synthetic ingots were prepared by first reacting the
corresponding pure elements and then subliming the compound in an
argon atmosphere as described in detail in Ref.
\onlinecite{Sherwin}.

The heat capacities were measured between 5 and 280K with a PPMS
system (Quantum Design, 6325 Lusk Boulevard, San Diego, CA.) as
described in Ref. \onlinecite{Serrano}. All natural samples measured
gave the same results within error bars.

\section{Experimental vs. Theoretical Results}
Figure \ref{fig3} shows the specific heat measured for our
$^{\rm{nat}}$Pb$^{\rm{nat}}$S samples for $T$ between 1.8 and 270K
and that reported in Ref. \onlinecite{Parkinson} for $T$ between 20
and 260 K. The solid line represents the results of our \textit{ab
initio} calculation. Whereas the two sets of experimental data agree
within error bars (the widths of the symbols in the plot), the
calculated curve falls slightly below the measurements, especially
around 80K. The sign of this deviation corresponds to calculated
phonon frequencies lying above the experimental ones, as is the case
for the TO modes in Fig. \ref{fig1}. Notice that at temperatures
above 280K $C_v$ tends to the Petit and Dulong value ($C_v$ $\sim$
49 Joule/mole K ) \cite{Petit}.

In order to be able to read the low temperature data of Fig.
\ref{fig3}, which according to Debye's law  should be proportional
to $T^3$, we display in Fig. 4 the quantity $C_v(T)/T^3$ . This
figure reveals that the region of validity of Debye's law is rather
small ($T <$ 3K). This follows from the flattening of the TA
dispersion relations with increasing \textbf{\textit{q}}. Figure
\ref{fig4} also shows that the calculated value of $C_v/T^3$ at its
maximum ($T_{max}$ $\sim$ 11K) is about 23\% lower than the measured
one. This discrepancy can be assigned to that of the calculated TO
frequencies (Fig. \ref{fig1}) which has already been attributed to
the lack of spin-orbit interaction in the band structure
calculations. We recall that in the case of bismuth a similar
discrepancy is found when the \textit{ab initio} calculations are
performed without spin-orbit coupling \cite{Diaz}.

Using Debye's theorem \cite{Debye}, Fig. \ref{fig4} can be recast
into a form which exhibits the temperature dependence of an
equivalent  $\theta_{\rm D}(T)$ (Fig. \ref{fig5}). This figure,
which includes calculated\cite{Elcombe} as well as experimental
data, reveals a value of $\theta_{\rm D}$ equal to 210 K in the $T
\rightarrow$ 0 limit. $\theta_{\rm D}(T)$ decreases to a minimum of
135 K at $\sim$ 20K.

$\theta_{\rm D}(T)$ saturates at $\theta_{\rm D}$ $\sim$ 230 K for
$T >$  230K. Note that in spite of the large scatter in our
experimental points, the agreement between theory and experiment is
rather good, except in the region around 75 K where a difference of
about 1\% appears. In order to get a feeling for the dependence of
the results of Fig. \ref{fig4} on isotopic masses, especially around
the maximum, we plot in Fig. \ref{fig6}a and \ref{fig6}b \textit{ab
initio} calculations for different sulfur (Fig. \ref{fig6}a) and
lead (Fig. \ref{fig6}b) isotopes. We note that a 2\% variation in
$M_{\rm Pb}$ results in a 3\% variation in $C_v /T^3$ whereas a 3\%
variation in $M_{\rm S}$ results in a 0.5\% variation of $C_v /T^3$.
This reveals a low sulfur content in the eigenvalues of the acoustic
modes as already discussed in connection with Fig. \ref{fig2}.

Figure \ref{fig7} displays the change in   $C_v /T^3$ measured and
calculated for a change of one a.m.u. in the mass of Pb in PbS.
Below 6K, the calculations (and also the measurements) yield
unreliable results. Correspondingly, the calculated curve has been
extrapolated to the value calculated for $T \rightarrow$ 0 using
Eqs. \ref{eq5} and the Debye limit for $C_v$ ( $C_v /T^3 \sim$
3$\mu$J/mol K$^4$). The maximum in both, experimental and calculated
data occurs for $T \sim$ 12 K. Note that the larger isotope mass
yields a larger value of  $C_v /T^3$.

Figure \ref{fig8} shows experimental and calculated results similar
to those in Fig. \ref{fig7} but obtained for a change of 2 a.m.u. in
$M_{\rm S}$. The calculated curve displays a maximum at 35 K. The
measured points scatter markedly around this temperature but, on the
average, they are also compatible with the existence of a maximum at
35 K. Below $\sim$10K both sets of data become unreliable. We have
also extrapolated the calculated curve to join the value predicted
with Eqs. \ref{eq5} and the Debye law.

\section{Discussion}
\begin{figure}[tbph]
\includegraphics[width=8cm ]{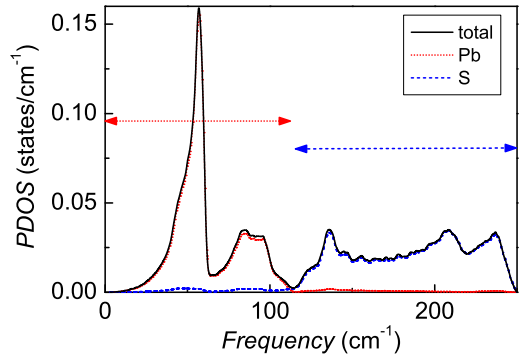}
\caption{(color online) Phonon density of states (PDOS) of PbS
(total, black solid line) obtained from the dispersion relations of
Fig. \ref{fig1} and its components projected on the Pb and S atoms
(red dots and blue dashes). The horizontal arrows indicate the
regions of of predominant Pb and S contributions, respectively. For
frequencies below (above) 120 cm$^{-1}$ the sulfur (lead) component
is extremely small (less then 1\%).} \label{fig2}
\end{figure}

\begin{figure}[tbph]
\includegraphics[width=8cm ]{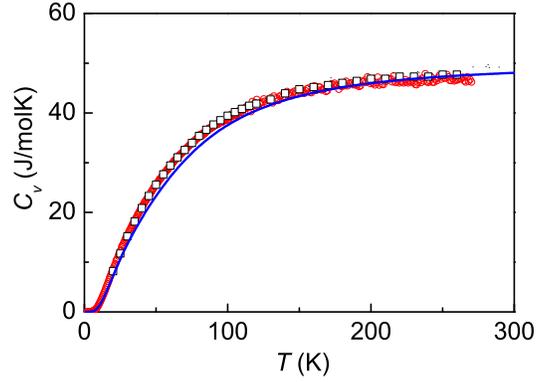}
\caption{ (color online) Experimental molar specific heat of PbS
(galena, natural isotope composition of Pb and S, respectively (red
circles). Also shown is the specific heat calculated from the PDOS
of Fig. \ref{fig2} (blue solid line) and the data by Parkinson
(galena, Ref. \onlinecite{Parkinson}).} \label{fig3}
\end{figure}

\begin{figure}[tbph]
\includegraphics[width=8cm ]{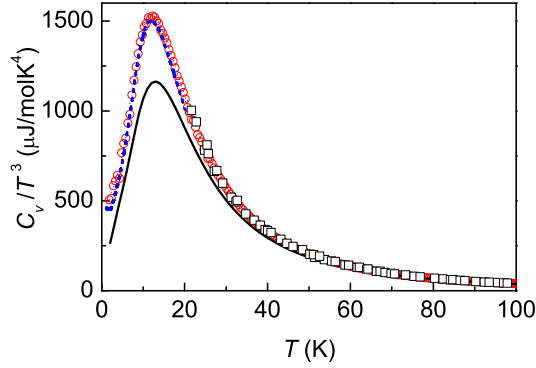}
\caption{(color online) Temperature dependence of $C_v/T^3$ as
obtained from the ABINIT calculation (lower (black) solid curve) and
measured by Lykov and Chernik (Ref. \onlinecite{Lykov}) (upper
dashed (blue) curve), by Parkinson (Ref. \onlinecite{Parkinson})
(black squares), and by us (red circles). The difference between the
measurements and the ABINIT calculations is attributed to the lack
of spin-orbit interaction in the latter.} \label{fig4}
\end{figure}

\begin{figure}[tbph]
\includegraphics[width=8cm ]{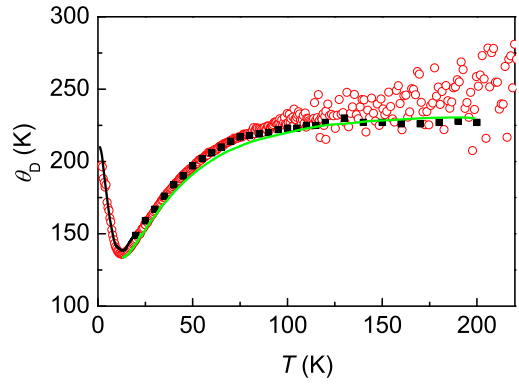}
\caption{(color online) Effective Debye temperature $\theta_{\rm
D}(T)$ (see text) of PbS. Our data for PbS (natural isotope
composition for Pb and S, respectively, are displayed by (red)
circles. Parkinson's data (Ref. \onlinecite{Parkinson}) by (black)
squares, the data reported by Lykov and Chernik (Ref.
\onlinecite{Lykov}) are represented by the (black) solid line. The
(green) solid curve represents the results of semiempirical
calculations by Elcombe (Ref. \onlinecite{Elcombe}).} \label{fig5}
\end{figure}

\begin{figure}[tbph]
\includegraphics[width=10cm ]{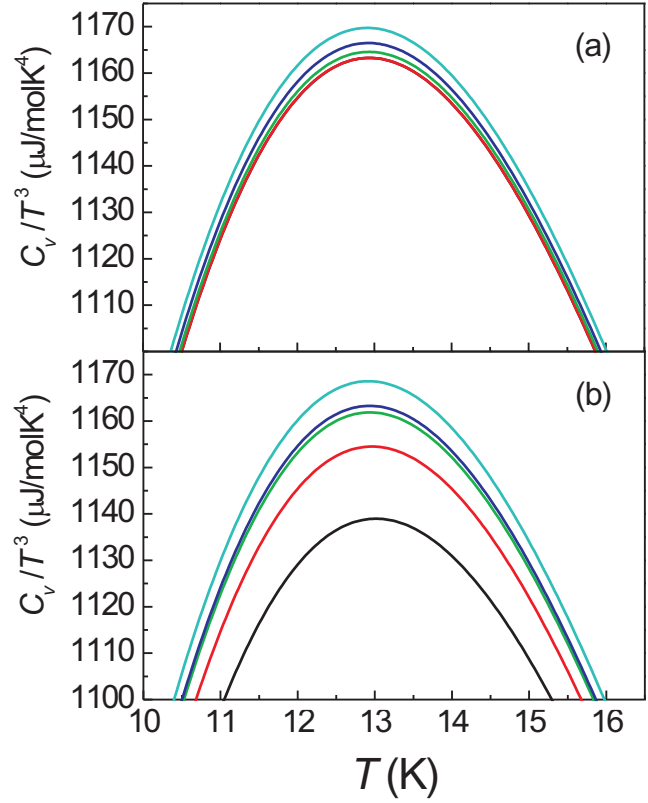}
\caption{(color online) ABINIT calculations of $C_v/T^3$ for PbS
with various isotope compositions. (a) $^{\rm{nat}}$Pb$^{\rm{n}}$S
with (from top to bottom) n=36, 34, 33, nat, 32 a.m.u. (note that
the curves for $^{\rm{nat}}$Pb$^{\rm{nat}}$S and
$^{\rm{nat}}$Pb$^{\rm{32}}$S lie on top of each other). The
superscript nat refers to the natural isotope abbundance of S which
corresponds to an atomic mass of 32.06 a.m.u. (b)
$^{\rm{m}}$Pb$^{\rm{nat}}$S with (from top to bottom) m=208, nat,
207, 206, 204 a.m.u. The superscript nat refers to the natural
isotope abbundance of Pb which corresponds to an atomic mass of
207.21 a.m.u. The isotope effects are much smaller in (a) than in
(b), as corresponds to the low sulfur content of the lower frequency
eigenvectors.} \label{fig6}
\end{figure}

As shown in Figs. \ref{fig3} and \ref{fig4} our calculations, based
on phonon dispersion relations obtained from \textit{ab initio}
electronic band structures, reproduce reasonably well the
experimental results. The discrepancy which appears at the maximum
of Fig. \ref{fig4} can be attributed to the lack of spin-orbit
interaction in the otherwise relativistic Hamiltonian used for the
\textit{ab initio} electronic calculations. The peak in the $\Delta
C_v/T^3$ of Fig. \ref{fig7} occurs at 12 K, a temperature about
one-third that of the peak seen in Fig. \ref{fig8}. The ratio of
both temperatures approximately equals the square root of the
corresponding masses. This fact allows us to assign the peak in Fig.
\ref{fig7} to vibrations of lead atoms and that in Fig. \ref{fig8}
to those of sulfur atoms, in agreement with the assignment of the
structures in the PDOS of Fig. \ref{fig2} below and above 120
cm$^{-1}$. The Pb-like PDOS of Fig. \ref{fig2} shows a sharp peak at
55 cm$^{-1}$ ($\sim$ 80K in temperature units) which has been
discussed in Sect. III. The S-like PDOS spreads over a broad band
with a peak at 133 cm$^{-1}$ ($\sim$ 196 K). The ratio of the
temperatures of these two peaks to those of the maxima in Figs.
\ref{fig7}and \ref{fig8} is $\sim$6. By modeling the two peaks in
the PDOS with single harmonic oscillators (Einstein model) it is
possible to calculate peaks in $\Delta C_v/T^3$ at temperatures
close to those displayed in Figs. \ref{fig7} and \ref{fig8}. For
simplicity, let us consider an Einstein oscillator with a frequency
equivalent to 100K which is inversely proportional to the square
root of its mass. The logarithmic derivative of the corresponding
specific heat with respect to the mass is

\begin{equation}
\frac{d (C_v/T^3)}{d ln M}\,\sim\,\frac{0.5}{100 T^2}\,\frac{d^2}{d
T^2}\frac{1}{(e^{\frac{100}{T}}-1)} = -T^{-6}\,
e^{\frac{100}{T}}\,\frac{e^{\frac{100}{T}}(T - 50)-T
-50}{(e^{\frac{100}{T}-1})^3}. \label{eq6}
\end{equation}

Calculations based on Eq. (\ref{eq6}) are plotted in Fig.
\ref{fig9}, with the vertical axis in arbitrary units.

The ratio of the Einstein oscillator temperature $T_{\rm E}$ to that
of the peak in Fig. \ref{fig9} is 100/16=6.25 and is independent of
$T_{\rm E}$. This ratio agrees rather well with the value obtained
from Figs. 7 and 2 (6.25) for the derivative vs. $M_{\rm Pb}$ and
for that vs. $M_{\rm S}$ from Figs. \ref{fig8} and \ref{fig2} (6.1).
It is interesting to compare these ratios with those obtained for
two rather similar materials: ZnO (Ref. \onlinecite{Serrano}) and
GaN (Ref. \onlinecite{Kremer}). For ZnO one obtains the ratio 7.3
for the $M_{\rm Zn}$ derivative and 5.1 for the $M_{\rm O}$
derivative, whereas in the case of GaN one finds 6.5 for the
$M_{\rm{Ga}}$ derivative and 3.5 for the $M_{\rm N}$ derivative.

We have already mentioned in Sec. II that for monatomic crystals the
mass derivative of the specific heat can be obtained from the
corresponding temperature derivative by using Eq. (\ref{eq4})
\cite{Sanati}. This equation cannot be used for crystals containing
several different atoms per primitive cell such as PbS, a fact which
becomes obvious when one considers that there is only one
temperature derivative but two (or more) independent mass
derivatives. It was shown, however, in Ref. \onlinecite{Kremer} (for
GaN) and Ref. \onlinecite{Serrano} (for ZnO) that the sum of the
corresponding cation and anion mass derivatives equals the result
obtained with Eq.(\ref{eq4}) from the temperature derivative. Figure
\ref{fig10} demonstrates that the same property applies to PbS. This
figure includes the results obtained by replacing into
Eq.(\ref{eq4}) the experimental as well as the \textit{ab initio}
calculated temperature derivatives. The figure also includes the two
appropriate mass derivatives and their sum (all from the \textit{ab
initio} calculations). Some difference appears between the latter
and the curve obtained from the measured temperature derivative near
the maximum at about 20K. We believe that their origin lies in
experimental and computational errors at these rather low
temperatures and also in the neglect of spin-orbit coupling in the
\textit{ab initio} calculations.

\section{Conclusions}
\begin{figure}[tbph]
\includegraphics[width=8cm ]{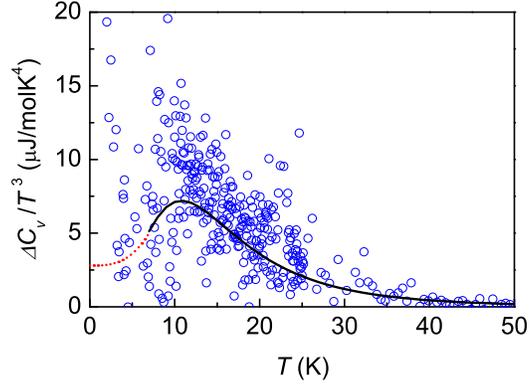}
\caption{(color online) $\Delta C_v/T^3$ difference for two samples
with natural sulfur and two different lead isotopic compositions
(natural and $M_{\rm Pb}$ = 208 a.m.u.). The (blue) circles
represent experimental points, the (black) solid curve calculations
which are not reliable below $\sim$ 6K. The (red) dotted curve
represents an extrapolation of the calculated curve for $T
\rightarrow$ 0 according to Eq. \ref{eq5}.} \label{fig7}
\end{figure}

\begin{figure}[tbph]
\includegraphics[width=8cm ]{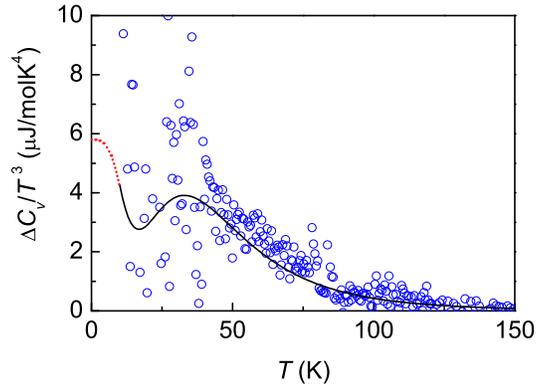}
\caption{(color online) $\Delta C_v/T^3$ difference for two samples
with natural Pb and two different sulfur isotopic compositions
(natural and $M_{\rm S}$ = 34 a.m.u.). The (blue) circles represent
experimental points, the (black) solid curve calculations which are
not reliable below $\sim$ 6K. The (red) dotted curve is an
extrapolation of the calculated curve for $T \rightarrow$ 0 using
Eq. \ref{eq5}.} \label{fig8}
\end{figure}

\begin{figure}[tbph]
\includegraphics[width=8cm ]{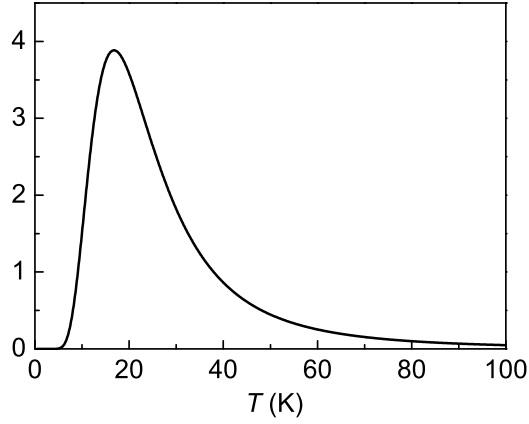}
\caption{Einstein oscillator model for the isotope effects of Figs.
\ref{fig7} and \ref{fig8}.  The curve was obtained with Eq.
(\ref{eq6}) which assumes, for simplicity, an Einstein temperature
of 100K. The temperature of the peak is 16.5K. The ratio of the
Einstein to the peak temperature is, $\approx$6, independent of the
Einstein temperature.} \label{fig9}
\end{figure}

\begin{figure}[tbph]
\includegraphics[width=8cm ]{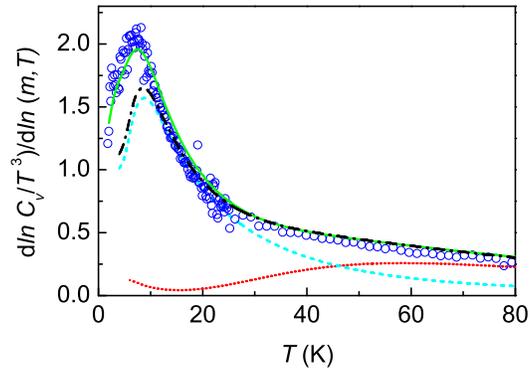}
\caption{ (color online) The (blue) circles represent the derivative
with respect to temperature, $d ln (C_v/T^3)/d ln T$ of the specific
heat of of PbS (galena). The data points were smoothed by averaging
five neighbor data points on either side (adjacent averages). The
(green) solid curve shows the r.h.s. of Eq. (\ref{eq4}), the
(magenta) dashed and (red) dotted curves the derivatives with
respect to the mass, $d ln (C_v/T^3)/d ln M$, with $M$ being the
mass of either lead or sulphur, respectively. The (black)
dashed-dotted solid curve is the sum of these derivatives (l.h.s. of
Eq. (\ref{eq4})). Note that above 40K the (green) solid and the
(black) dashed-dotted curves lie on top of each other.}
\label{fig10}
\end{figure}

We have investigated experimentally and theoretically (\textit{ab
initio}) the specific heat of PbS and its derivatives with respect
to the masses of the two constituents. Reasonable agreement is found
between experiment and calculations, quantitative differences having
been attributed to the absence of spin-orbit interaction in the
latter.  The mass derivatives vs. $T$ exhibit maxima at about 12 K
for lead and 33 K for sulfur. The large difference in the masses of
both constituents allows us to separate these maxima and to relate
them to corresponding maxima in the PDOS. The frequency of the
latter is about six times larger than that of the former: a simple
model, based on a single Einstein oscillator, has been proposed to
account for this factor. Qualitatively similar ratios have been
found in recent years for ZnO and GaN, two materials for which the
ratio of cation to anion mass is not as large as for PbS. Note that
the  Einstein oscillator curve of  Fig.\ref{fig9}     can be used,
by adjusting the oscillator frequency and the vertical scale, to
give a rather good representation of the experimental data  (and the
calculations) of Figs. \ref{fig7} and \ref{fig8}.

By inverting the integral equation relating the specific heat and
its derivatives to the components of the phonon eigenvectors on the
two (or more) constituent atoms, it should be possible to obtain the
corresponding projected densities of states from experimental data.
Because of the large error bars in some of our measured points,
however, we have not implemented this procedure. \textit{ab initio}
calculations of the projected PDOS have led us to the conclusion,
that below 120 cm$^{-1}$, the eigenvectors are to a large degree
lead -like, whereas above this frequency they are almost completely
sulfur-like.

In order to remove remaining discrepancies between theory and
\textit{ab initio} calculations it should be interesting to modify
the ABINIT code so as to include spin-orbit coupling in the phonon
calculations. Spin-orbit coupling is only included, in the ABINIT
version available to us for calculating the specific heat, for
monatomic materials such as bismuth \cite{Diaz,Murray}. A procedure
to include spin-orbit splitting in lattice dynamical calculations
with ABINIT would be highly desirable in order to ascertain the
contributions in the phonon dispersion and thermodynamical
properties in lead chalcogenides, among other multinary
semiconductors.

\begin{acknowledgments}
We thank L. Wirtz and O. Kilian for very fruitful discussions and
for making available to us unpublished results on the phonon
dispersion of lead chalcogenides.

\end{acknowledgments}

\end{document}